\documentclass[aps,prb,twocolumn]{revtex4}
\usepackage{graphicx}
\usepackage{epsfig}
\usepackage{amsmath}
\usepackage{amssymb}
\usepackage{amsfonts}%
\newcommand{\pup}{\hat {\bf p}}
\begin{document} 
\title{Quasiclassical approach to the spin-Hall effect in the
two-dimensional electron gas}  
\author{Roberto Raimondi}
\author{Cosimo Gorini}
\affiliation{Dipartimento di Fisica "E. Amaldi", Universit\`a di Roma Tre, Via della Vasca Navale 84, 00146 Roma, Italy}
\author{Peter Schwab}
\author{Michael Dzierzawa}
\affiliation{Institut f\"ur Physik, Universit\"at Augsburg, 86135 Augsburg, Germany}

\begin{abstract} 
We study the spin-charge coupled transport in a two-dimensional
electron system using the method of quasiclassical ($\xi$-integrated)
Green's functions. In particular we derive the  Eilenberger equation in the presence of a generic spin-orbit field. 
The method allows us to study spin and charge transport from ballistic to diffusive regimes and 
continuity equations for spin and charge are automatically incorporated.
In the clean limit we establish the connection
between the spin-Hall conductivity and the Berry phase in momentum
space.
For finite systems we solve the Eilenberger equation numerically   
for the special case of the Rashba spin-orbit coupling and a
two-terminal geometry. In particular, we  calculate explicitly the spin-Hall induced
spin polarization in the corners, predicted by Mishchenko et al.
[\onlinecite{mishchenko2004}]. Furthermore
we find universal spin currents in the short-time dynamics after
switching on the voltage across the sample, and calculate the
corresponding spin-Hall polarization at the edges. Where available,
we find perfect agreement with analytical results. 
\end{abstract} 
\pacs{PACS numbers: } 

\date{\today} 
\maketitle 
\section{Introduction}
In the presence of spin-orbit coupling spin currents and spin
polarization can be generated as a response to electric fields
[\onlinecite{dyakonov1971,edelstein1990,hirsch1999,murakami2003,sinova2004}].
Recently the spin-Hall effect, i.e. a spin current that flows
perpendicular to an applied electric field has been observed
experimentally in electron doped semiconductors and in a
two-dimensional hole
system [\onlinecite{kato2004,wunderlich2005}].
Theoretically one may distinguish the extrinsic from the intrinsic spin-Hall
effect, depending on whether spin-orbit coupling arises due to
scattering by
impurities or from the intrinsic band structure of the samples.
The intrinsic spin-Hall effect has first been studied for holes in p-type semiconductors in [\onlinecite{murakami2003}]
and for electrons in n-type semiconductors in [\onlinecite{sinova2004}].
In both cases, the striking result is the independence of the
spin-Hall conductivity $\sigma_{sH}$ of the strength
of the spin-orbit coupling, at least when disorder effects are ignored.
Sinova {et al.} [\onlinecite{sinova2004}] found in the two-dimensional
electron system with Rashba spin-orbit coupling a universal value
for the spin-Hall conductivity, $\sigma_{sH} = e/8 \pi$.
Soon it was realized that the universal spin-Hall conductivity 
still exists in the presence of a Dresselhaus term
[\onlinecite{sinitsyn2004,shen2004}], and even in the presence of a weak
in-plane magnetic field [\onlinecite{chang2005}].
It was pointed out that this may be related to a
Berry phase in momentum space, i.e.\ the winding number of the
spin-orbit field when going once around the Fermi surface.
Shytov et al.\ [\onlinecite{shytov2005}] showed such a connection
explicitly in the specific case where the modulus of the spin-orbit field remains
constant on the Fermi surface.

Clearly it is an important question to ask how 
$\sigma_{sH}$ depends on disorder.
For the Rashba model the effect is quite dramatic, namely an
arbitrarily weak amount of disorder fully suppresses the spin-Hall
conductivity
[\onlinecite{inoue2004,mishchenko2004,khaetskii2004,raimondi2005,dimitrova2005,chalaev2005,nomura2005}].
Meanwhile it is understood that this surprising result is a special property
of the Rashba Hamiltonian and is related to the
linear-in-momentum spin-orbit field: The time derivative of the total
spin is proportional to the spin current, so that in a steady state
both quantities are zero [\onlinecite{rashba2004,dimitrova2005}].
In the case of a more general spin-orbit field, a finite spin-Hall
effect has been reported even in the presence of disorder
[\onlinecite{murakami2004,malshukov2005,nomura2005,shytov2005,khaetskii2005}].

With the conventional definition of the spin current, given by the
anticommutator of the velocity operator and the Pauli matrices, the
spin current is not conserved. 
Hence it is not
automatically guaranteed that a current in the bulk induces a
polarization at the edges of the sample.
Furthermore the spin current is not directly accessible
experimentally; the measurable quantity is the spin polarization
instead.
Therefore it is of interest to study directly the electric field
induced spin-density.
A strategy followed by some authors is to
discretize the Rashba Hamiltonian in terms of a tight-binding model
which is then studied near the band edge. In this way the spin-Hall
induced spin accumulation in  
systems with linear dimensions of several tens of
the Fermi wavelength have been studied
[\onlinecite{nikolic2005,onada2005}].
Macroscopic systems are conveniently described in
terms of semiclassical kinetic equations 
[\onlinecite{burkov2004,mishchenko2004,khaetskii2004,malshukov2005,shytov2005,khaetskii2005}],
and in particular in terms of diffusion equations describing the
coupled dynamics of spin and charge degrees of freedom.
In clean systems, however, the spin relaxation length
can become comparable to the elastic mean free path. In this situation
the spin dynamics is not diffusive, and
one has to go beyond the diffusive approximation.
In what follows we present a theory of the spin-Hall effect in terms
of quasiclassical Green's functions, which covers the full range from
clean systems to the diffusive limit.
The first step is the derivation of the equation-of-motion, the Eilenberger equation. 
From there we derive the continuity equation for charge and spin, and
obtain explicit expressions for the current densities.
For the Rashba
model it follows that no spin current can flow in a time
and space independent situation.
In the clean limit we find a
solution of the equation-of-motion which corresponds to the universal spin-Hall effect.
In the general case of disordered finite systems we solve the
equation-of-motion numerically.

\section{The Eilenberger equation}

We start from the Hamiltonian
\begin{equation}
H = \frac{p^2}{2m} + {\bf b} \cdot {\boldsymbol \sigma }
\end{equation}
where ${\bf b}$ is the internal magnetic field due to the spin-orbit
coupling
and ${\boldsymbol \sigma}$ is the vector of Pauli matrices.
In the Rashba model for example 
${\bf b } = \alpha {\bf p} \times {\bf e}_z $.
For a spin $1/2$ particle one can  write the spectral decomposition of the Hamiltonian 
in the form 
\begin{equation}
H = 
\epsilon_+ \,  |\! + \! \rangle \langle \! + \! | + 
\epsilon_- \,  |\! - \! \rangle \langle \! - \! | 
\end{equation}
where $\epsilon_\pm = p^2/2m \pm |{\bf b }| $ are the eigenenergies 
corresponding to the projectors
\begin{equation}
 | \pm \rangle \langle \pm | =  
\frac{1}{2}\left( 1 \pm \hat {\bf  b} \cdot {\boldsymbol \sigma }
\right)
\end{equation}
where $\hat {\bf  b}$ is the unit vector in the ${\bf b}$ direction.
We write  Green's functions in Wigner coordinates,
$G= G({\bf p}, {\bf x} )$, where ${\bf p}$ 
is
the Fourier transform of the relative coordinate 
and $\bf x $ is the 
center-of-mass coordinate.
For the Green functions we make the ansatz
\begin{equation}
\check G = 
\left( \begin{array}{cc} 
G^R & G^K \\
0   & G^A 
\end{array}
\right)
= \frac{1}{2} \left\{ 
\left( \begin{array}{cc} 
G^R_0 & 0 \\
0   & -G^A_0 
\end{array}
\right),
\left( \begin{array}{cc} 
\tilde g^R & \tilde g^K \\
0   & \tilde g^A 
\end{array}
\right)
\right\}
\end{equation}
where 
the curly brackets denote the anticommutator.
$G^{R,A}_0$ are  retarded and advanced Green's functions in the absence of external
perturbations,
\begin{equation}
G_0^{R(A)} = \frac{1 }{\epsilon+ \mu- {p^2/2m} - {\bf b} \cdot {\boldsymbol \sigma } -
\Sigma^{R(A)}  }
,\end{equation}
and $\Sigma^{R(A)}$ are the retarded and advanced self-energies which will be specified below. 
The ansatz guarantees that in equilibrium the matrix of Green's 
functions with small letters is
\begin{equation}
\check{\tilde g } = \left( \begin{array}{cc}
1 & 2 \tanh(\epsilon/2T) \\
0   & -1
\end{array}
\right)
.\end{equation}
The main assumption for the following is that  we can determine
$\check{\tilde g }$ such that it does not depend on the modulus of the
momentum ${\bf p}$ but only on the direction $\hat {\bf p}$.
Under this condition $\check{\tilde g }$ is directly related to the
$\xi$-integrated Green function which we denote by $\check{ g }$,
\begin{equation}
\check{ g } = \frac{\rm i}{\pi} \int {\rm d } \xi \, \check G,
\quad \xi = p^2/2m - \mu
.\end{equation}
For convenience we suppressed in the equations above spin and time 
arguments of the Green function, 
$\check g = \check g_{t_1 s_1, t_2 s_2}(\hat {\bf p};{\bf x})$.
In some cases Wigner coordinates for the time arguments are more
convenient,
$\check g \to \check g_{s_1 s_2}(\hat {\bf p}, \epsilon; {\bf  x}, t) $.

We evaluate the $\xi$-integral explicitly in the limit 
where $|\bf b|$ is small compared to the Fermi energy.
Since the main contributions to the $\xi$-integral are from the region
near zero, it is justified to expand ${\bf b}$ for small $\xi$,
${\bf b} \approx {\bf b}_0 + \xi \partial_{ \xi} {\bf b}_0 $, with the
final result
\begin{eqnarray} \label{eq8}
\check g & \approx    &  
\frac{1}{2} \left\{ 1 - \partial_\xi {\bf b}_0 \cdot {\boldsymbol \sigma }
,\check { \tilde g}  \right\} \\
\check {\tilde g } & \approx  & \frac{1}{2} \left\{ 1 + \partial_\xi
{\bf b}_0 \cdot {\boldsymbol \sigma }
,\check { g}  \right\}.
\end{eqnarray}
In the equation of motion we will also have to evaluate integrals of
a function of $\bf  p$ and a Green's function. Assuming again that
$|{\bf b}| \ll \epsilon_F$ we find
\begin{equation}
\frac{\rm i}{\pi } \int {\rm d } \xi \, f({\bf p}) \,  \check G
 \approx  
f({\bf p}_+)\check g_+ + f({\bf p}_- ) \check g_- 
\end{equation}
where ${\bf p}_\pm $ is the Fermi momentum in the $\pm$-subband
including corrections due to the internal field, 
$|{\bf p}_\pm |\approx  p_F \mp |{\bf b}|/v_F $, and
\begin{equation} \label{eq11}
 \check g_\pm = \frac{1}{2} 
 \left\{
\frac{1}{2}  \pm \frac{1}{2} \hat{\bf b}_0 \cdot {\boldsymbol \sigma }
, \check g  \right\}, \quad \check g = \check g_+ + \check g_-
.\end{equation}

Following the conventional procedure \cite{rammer1986} we derive now
the equation-of-motion for $\check g$.
From the Dyson equation and
after a gradient expansion the equation-of-motion for the Green
function $\check G$ reads
\begin{eqnarray}
&& \partial_t  \check G +
\frac{1}{2} \left\{  
\frac{\bf p}{m} + \frac{\partial}{\partial {\bf p } }({\bf b}\cdot
{\boldsymbol  \sigma}),
\frac{\partial}{\partial \bf x} \check G 
\right\}
+ {\rm i} 
\left[
{\bf b}\cdot {\boldsymbol \sigma}, \check G 
\right] \nonumber\\[1mm]
&& \label{eq12} =
- {\rm i} [\check \Sigma,  \check G ]. 
\end{eqnarray}
The Boltzmann equation or Boltzmann-like kinetic equations are
obtained by either integrating (\ref{eq12}) over energy $\epsilon$ or
over $\xi$, see [\onlinecite{rammer1986}]. 
Ref. [\onlinecite{shytov2005}] for instance follows the first route,
whereas we integrate over $\xi$.
Retaining terms up to first order in
$|{\bf b}|/\epsilon_F$ leads to an Eilenberger equation of the form
\begin{eqnarray}
\sum_{\nu = \pm }\big(
  \partial_{t} \check g_\nu  &+ &
  \frac{1}{2} \left\{
    \frac{\bf p_\nu}{m}  
   +\frac{\partial}{\partial \bf p}({\bf b}_\nu \cdot {\boldsymbol  \sigma}),
    \frac{\partial}{\partial {\bf x} }\check g_\nu \right\} \nonumber \\
&+&  {\rm i } [{\bf b}_\nu \cdot {\boldsymbol \sigma}, \check
g_\nu ]\big)
    \nonumber \\[1mm]
&=& - {\rm i} \left[ \check \Sigma , \check g \right].
\label{eqEilenberger}
\end{eqnarray}
In the entire article we will take the self-energy as
$\check \Sigma =-\rm i \langle \check g \rangle  /2\tau$ , which corresponds to $s$-wave impurity scattering in
the Born approximation;  $\langle  \dots \rangle$ denotes the angular
average over $\hat { \bf p } $.

To check the consistency of the equation we study at first its retarded
component in order to verify that $\tilde g^R = 1$ solves the generalized
Eilenberger equation.
From Eq.~(\ref{eq8}) we find that
$g^R = 1- \partial_\xi ({\bf b}_0 \cdot {\boldsymbol \sigma } )$ and 
using (\ref{eq11}) we arrive at
\begin{equation}
g^R_\pm = ( 1 \mp \partial_\xi b )\left(  \frac{1}{2}  \pm 
\frac{1}{2} \hat {\bf b}_\pm \cdot {\boldsymbol \sigma}  \right)
.\end{equation}
Apparently both the commutators on the left and on the right hand side
of the Eilenberger equation are zero, at least to first order in the
small parameter $\partial_\xi {\bf b}_0$.
Similar arguments may also be used to verify that 
the equilibrium Keldysh component of the Green function,
$g^K  = \tanh (\epsilon/2 T )( g^R - g^A )$, solves the equation of
motion.

In the appendices we demonstrate how the frequency dependent
spin-Hall conductivity and the equation-of-motion in the diffusive limit
can be obtained from Eq.~(\ref{eqEilenberger}). For the Rashba model
our results agree with Ref.~[\onlinecite{mishchenko2004}].

\section{Continuity equation -- vanishing spin-Hall current}
Equipped with the Eilenberger equation it is not difficult to see that
for a spin-orbit field of the Rashba or linear Dresselhaus model the
spin-Hall conductivity is zero. 
The argument is analogous to that of Ref.~[\onlinecite{dimitrova2005}] and makes use of
the continuity equation.

When taking the angular average of the Eilenberger equation
(\ref{eqEilenberger}), the
term on the right hand side vanishes and we are left with a set of 
continuity equations for the charge and spin components of the Green
function.
With 
$\check g_{ss'} = \check g_0 \delta_{ss'} + \check  {\bf  g } \cdot {\boldsymbol \sigma}_{ss'}$
the equations read
\begin{eqnarray}
\partial_t \langle \check g_0 \rangle + \partial_{\bf x} \cdot \check {\bf J}_{c} &=&0 \\
\label{eq16}
\partial_t \langle \check g_x \rangle + 
\partial_{\bf x} \cdot\check {\bf J}_{s}^x&=&2
\sum_{\nu = \pm }  \langle {\bf b}_\nu \times {\bf \check g_\nu} \rangle_x \\
\label{eq17}
\partial_t \langle \check g_y \rangle + 
\partial_{\bf x } \cdot \check {\bf J}_{s}^y&=&2
\sum_{\nu = \pm }  \langle {\bf b}_\nu \times {\bf \check g_\nu} \rangle_y \\
\partial_t \langle \check g_z \rangle + 
\partial_{\bf x}\cdot \check {\bf J}_{s}^z  &=&2
\sum_{\nu = \pm }  \langle {\bf b}_\nu \times {\bf \check g_\nu} \rangle_z 
\end{eqnarray}
with 
\begin{equation}
\check {\bf J}_{c,s} = \sum_{\nu=\pm} \left\langle   
\frac{1}{2} \left\{
\frac{{\bf p}_\nu} {m} + \frac{\partial}{\partial {\bf p}}({\bf b}_\nu
\cdot \boldsymbol{\sigma}), \,  \check g_\nu
\right\}
\right\rangle_{c,s}.
\end{equation}
The densities and currents are related to the Keldysh components of
$\langle \check g \rangle $ and of $\check {\bf J}_{c,s}$ integrated
over $\epsilon$.
Explicitly the particle and spin current densities are given by
\begin{eqnarray}
{\bf j}_c({\bf x}, t ) & = & - \pi N_0 \int \frac{{\rm d} \epsilon }{2
\pi }{\bf J}^K_{c}(\epsilon; {\bf x},t ) \\
{\bf j}_s^i({\bf x}, t) &= & - \frac{1}{2} \pi N_0  \int \frac{{\rm
d}\epsilon}{2 \pi }{\bf J}^{K i}_{s}(\epsilon; {\bf x}, t)
\end{eqnarray}
with $N_0 = m/ 2 \pi $ being the density of states of the
two-dimensional electron gas.
In the the absence of spin-orbit coupling (${\bf b} = 0$) one recovers the
well known expressions
\begin{eqnarray}
{\bf j}_c   ({\bf x}, t)  & =  &- \frac{1}{2} N_0 \int {\rm d } \epsilon  \langle
{\bf v}_F g_0^K  \rangle   \\
{\bf j}_s^i ({\bf x}, t)  & =  &-\frac{1}{4}  N_0 \int{\rm d } \epsilon\langle  {\bf v}_F g_i^K \rangle.
\end{eqnarray}
In the presence of the field ${\bf b}$ the expressions are in general
more complex.
For the Rashba model, for example, the particle current is given by
the lengthy expression
\begin{eqnarray}
{\bf j}_c({\bf x}, t )& = & - \frac{1}{2} N_0 \int {\rm d} \epsilon  [
v_F   \langle \pup g_0^K \rangle  \cr
   && +  
           \alpha( \hat{\bf e}_z \times \langle {\bf g}^K \rangle
           - \langle  \pup ( \pup \cdot \hat{\bf e}_z \times {\bf g}^K
             \rangle ) ]
 .\end{eqnarray}

Finally let us consider the spin current with polarization in
$z$-direction for the model with both a Rashba and a Dresselhaus term for
which the spin-orbit field reads
\begin{equation}
\left(\begin{array}{c} b_x \\ b_y \\ b_z  \end{array} \right)
= 
\alpha 
\left( \begin{array}{c} p_y \\ -p_x  \\ 0  \end{array}  \right)
+ \beta 
\left( \begin{array}{c} p_x \\ - p_y \\ 0   \end{array}  \right)
.\end{equation}
Because the  field lies in the $x-y$-plane,
 ${\bf {\check J}}_s^z$ is simply given by
${\bf {\check J}}_s^z = \langle {\bf v}_F {\check g}_z  \rangle $. Besides the field is
just linear in $\bf p$. 
As a result we find that
the source term in
the continuity equations (\ref{eq16}) and (\ref{eq17}) can be expressed in terms of the spin current:
\begin{eqnarray}
(\ref{eq16}) 
&=& \hphantom{-} 
2 \langle b_{y,0} {\check g}_z \rangle
= -2 m \alpha {\check J}_{s,x}^z- 2m \beta {\check J}_{s,y}^z
\\
(\ref{eq17})
&=& - 2 \langle b_{x,0} {\check g}_z
\rangle = -2m \alpha {\check J}_{s,y}^z - 2 m \beta {\check J}_{s,x}^z.
\end{eqnarray}
In a stationary situation and for a spatially homogeneous system the
left hand side of the continuity equation is zero and this implies
a vanishing spin current, 
\begin{equation} j_{s, x}^z  = j_{s,y}^z = 0
.\end{equation}

\section{Clean limit -- universal spin-Hall currents}
\label{CleanLimit}
We consider now the Eilenberger equation in the clean limit, 
$\tau \rightarrow \infty$, and study the linear response to an
homogeneous electric
field. For a realistic system with at least weak disorder
this study still gives reliable results on short time scales, 
$t \ll \tau$.
Generally an electric field can be included in the
quasiclassical equations-of-motion by the substitution
$\partial_{\bf x} \to \partial_{\bf x} - |e| {\bf E}\partial_\epsilon$.
The Keldysh component of the linearized Eilenberger equation becomes
\begin{eqnarray}
&& \sum_{\nu= \pm } \big( \partial_t g_\nu^K - \frac{|e |}{m } {\bf E}  \cdot {\bf p}_\nu \partial_\epsilon
g_\nu^{K,\rm eq} \label{linearkeldysh} \\
&& -\frac{|e|}{2} \left\{ ( {\bf E}\cdot \partial_{\bf p}) ( { \bf b }_{\nu}
  \cdot {\boldsymbol \sigma } ) , \partial_\epsilon g_\nu^{K,\rm eq}
  \right\} 
 + {\rm i } [{\bf b}_\nu \cdot {\boldsymbol \sigma}, g_\nu^K ]
\big) = 0. \nonumber 
\end{eqnarray}
In the following we focus on the spin components of the equation.
Explicitly we get
\begin{eqnarray}
\partial_t g_x^K & = & \hphantom{-} 2 b_{y,0} g_z^K \cr 
&& +  
 |e| {\bf E} \cdot \left[{\bf P}{\hat b}_{x,0} -{\bf v}_F \partial_\xi
 b_{x,0} + \partial_{\bf p} b_{x,0} \right]  F_\epsilon \label{gx} \\
\partial_t g_y^K & = & -2 b_{x,0} g_z^K  \cr
&&+ 
 |e| {\bf E} \cdot \left[{\bf P} {\hat b}_{y,0} -{\bf v}_F \partial_\xi
 b_{y,0} +  \partial_{\bf p} b_{y,0}  \right]  F_\epsilon   \label{gy} \\
\partial_t g_z^K & = & 2( b_{x,0} g_y^K - b_{y,0} g_x^K ) \cr
 && + 2 (b_{x,0}\partial_\xi b_{y,0} - b_{y,0} \partial_\xi b_{x,0} ) g_0 . 
\end{eqnarray}
where for the sake of brevity ${\bf P}=  \sum_{\nu} \nu{\bf p}_{\nu}/2m$ and $F_\epsilon =2  \partial_\epsilon \tanh(\epsilon/2T)$.  
For the $g_z^K$ component one obtains
\begin{equation}
\frac{{\rm d}^2 g_z^K}{{\rm d}t^2}+4 b_0^2 g_z^K =   2F_\epsilon 
 |e| \left[ 
b_{x,0} (  {\bf E } \cdot \partial_{\bf p } ) b_{y,0}-
b_{y,0} (  {\bf E } \cdot \partial_{\bf p } ) b_{x,0}
\right] 
.\end{equation}
Notice that only the second of the two terms involving the electric field in Eq.(\ref{linearkeldysh}) remains
in the equation for the  $g_z^K$ component.
The solution of this differential equation is the sum of an
oscillating and a time independent term.
Due to the (undamped) oscillations it is clear that
a stationary solution is never reached so the arguments of the
previous section leading to vanishing spin-Hall current do not apply.
The time independent solution of the differential equation is related
to a zero-frequency spin current given by,  
\begin{equation} \label{eq45}
{\bf j}^z_s = - \frac{|e|}{4 \pi} \langle {\bf p}_F ( { \bf E } \cdot
\partial_{\bf p } ) \Psi  \rangle , \quad \tan \Psi = b_{y,0}/b_{x,0}
.\end{equation}
Notice that the spin current does not depend on the magnitude of the
field $\bf b$, but only on the variation of its direction when going
around the Fermi surface.
An even more explicit result is obtained when the spin-Hall
conductivity tensor is antisymmetric
\begin{eqnarray}
\sigma_{sH} &  =  &\frac{1}{2} (\sigma_{sH}^{y,x}-\sigma_{sH}^{x,y} )
\\&=&
-\frac{|e|}{8 \pi} \langle ( 
p_{Fy} \partial_{p_x} - 
p_{Fx} \partial_{p_y} ) \Psi \rangle \\ 
\label{eq48}
&=&
\frac{|e|}{8\pi} \oint \frac{ {\rm d } {\bf p } }{2 \pi} \cdot \partial_{\bf p} \Psi
,\end{eqnarray}
i.e.\ the spin-Hall conductivity is the universal number $|e|/8\pi$ times
the winding number of the internal  field $\bf b $ when going once around
the Fermi surface.

We notice that Eq.~(\ref{eq45}) is consistent with 
[\onlinecite{sinitsyn2004,shen2004}] and also with
[\onlinecite{chang2005}] where the spin-Hall conductivity ignoring disorder
has been calculated using the Kubo formula
for a Rashba-Dresselhaus system in the presence of
an in-plane magnetic field.
Eq.~(\ref{eq48}) which relates the spin-Hall conductivity with a
winding number, i.e. the Berry phase in momentum space, generalizes
the equivalent result of [\onlinecite{shytov2005}], where it has been assumed
that the modulus of ${\bf b}$ is constant on the Fermi surface.
\section{Finite and disordered systems -- Numerical results}
In this section we solve  Eq.~(\ref{eqEilenberger}) numerically
for the Rashba model. 
Compared to the diffusion equations
\cite{burkov2004,mishchenko2004,malshukov2005}, 
one advantage of our method is that 
we have access to length scales that
are shorter than the mean free path $l=v_F \tau$. This is crucial in weakly
disordered systems, $|{\bf b} | > 1/\tau$, where the characteristic
length scale for the spin polarization, the spin relaxation
length, is of the order of the mean free path.
Furthermore,  when considering time-dependent situations we can
study the time evolution on time scales which are shorter than the
scattering time $\tau$.

In the following we will consider a geometry as shown in
Fig.~\ref{fig1}:
\begin{figure}
 \centerline{\includegraphics[height=3.0cm]{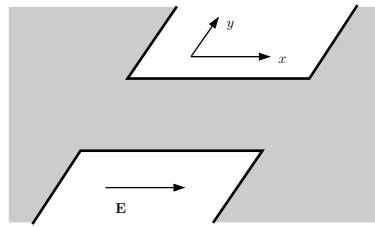}}
\caption{
\label{fig1}
The two-terminal geometry under consideration: A rectangular strip of a
two-dimensional electron gas is connected to two reservoirs. We assume
that the strip and the reservoirs are made of the same material, i.e.
the spin-orbit field exists also in the reservoirs.}
\end{figure}
A rectangular strip of length $L_x$
and width $L_y$, is connected to leads at $x=0$ and $x=L$.
At interfaces 
the Eilenberger equation has to be complemented with boundary
conditions \cite{zaitsev1984,shelankov1985,millis1988}.
The boundary condition between the strip and the leads is obtained
assuming that 
the leads are made of the same material as the strip, i.e.\ there is
no Fermi surface mismatch, and 
that both leads are in thermal equilibrium. 
For directions $\hat {\bf p}$ pointing into the strip
the Green function at the interface ($x=0,L$) reads
\begin{eqnarray} 
g^K(\hat {\bf p}_{\rm in}; {\bf x})\Big|_{x=0,L}  \! \! \!  & =
&g^K_{\rm eq}(\hat {\bf p}_{\rm in};{\bf x})\Big|_{{\bf x} \text{ in the lead}}  \\ 
\label{eq50}
&= &
\tanh\left(\frac{\epsilon \pm |e| V /2}{2 T }\right) (g^R-g^A)
,\end{eqnarray}
where $V$ is the applied voltage.
At the boundary with an insulator the Green functions for
in and outgoing directions are related via a surface scattering
matrix,
\begin{equation}
g^K(\hat {\bf p}_{\rm out } )  =  S g^K( \hat {\bf p }_{\rm in }) S^+
.\end{equation}
Generally, the $S$-matrix can be calculated  
by solving the quantum mechanical surface scattering problem.
For simplicity we assume specular scattering and assume that boundary
scattering does not induce transitions between the two 
spin-orbit subbands, so that 
\begin{equation}
| {\bf k}_{\rm in } \pm \rangle  \to  \exp({\pm {\rm i } \vartheta}) |
{\bf k}_{\rm out} \pm \rangle, \end{equation}
as it is found for smooth confining potentials
[\onlinecite{silvestrov2005}]. 
Explicitly the surface $S$-matrix for the Rashba Hamiltonian is 
\begin{equation}
S = {\rm e }^{ {\rm i } \varphi} 
\left( \begin{array}{cc} 
{\rm e }^{ {\rm i }  \varphi} \cos \vartheta & - \sin \vartheta  \\
                          \sin \vartheta  & {\rm e}^{- {\rm i }\varphi
                          } \cos \vartheta 
\end{array} \right)
,\end{equation}
where the angle $\varphi$ characterizes the ingoing
direction, $\hat{\bf  p}_{\rm in} =( \cos \varphi, \sin \varphi ) $.
Our numerical results are obtained assuming that the relative phase-shift
in the scattering for the two subbands is negligible, i.e.\  $\vartheta = 0$.

Finally, to integrate the equation of motion numerically we have to discretize the space coordinate ${\bf x}$ and the Fermi
surface. In dirty systems $g^K(\hat {\bf p})$ is nearly isotropic, so it
is clear that a few discrete points $\hat {\bf p}_i $ on the Fermi
surface are sufficient. 
In clean systems this is not {\sl a priori} evident, but numerical
tests show that even in this case  
convergence is reached quickly. Typically we describe the Fermi
surface with a set of twenty to forty $\hat {\bf p}_i $. 

\begin{figure}
\includegraphics[width=0.48\textwidth]{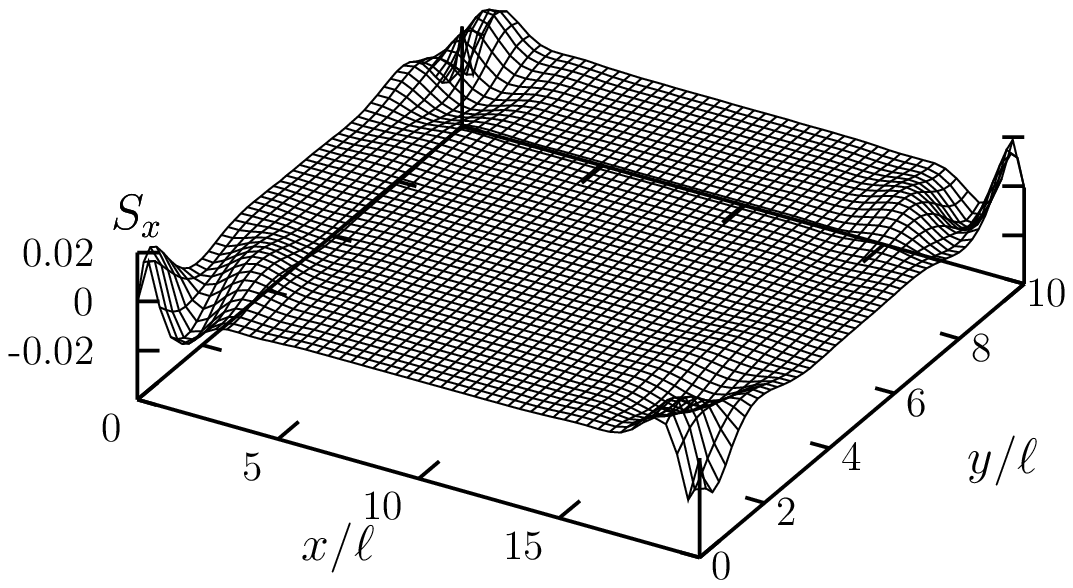} 
\includegraphics[width=0.48\textwidth]{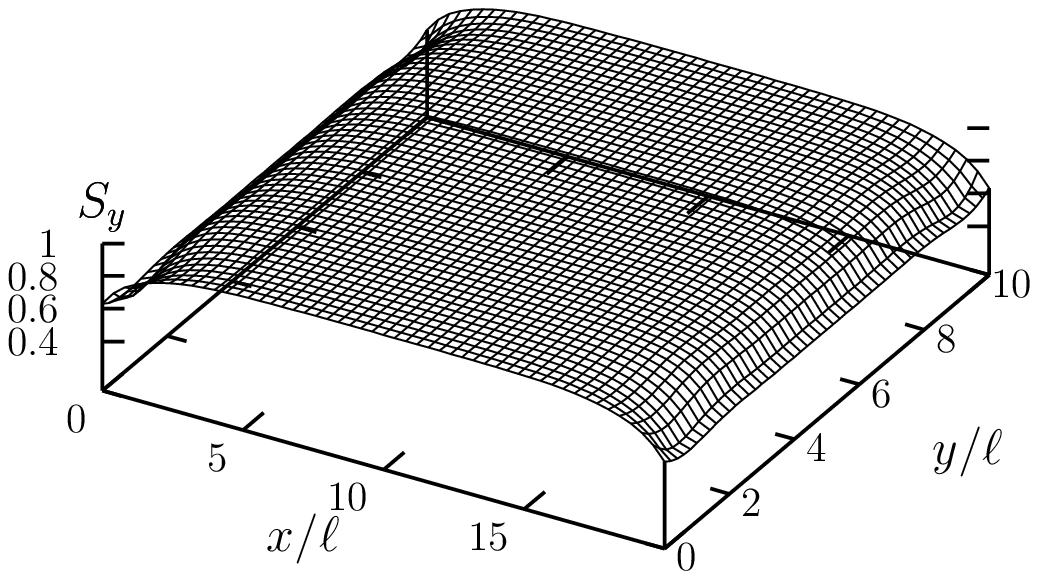} 
\includegraphics[width=0.48\textwidth]{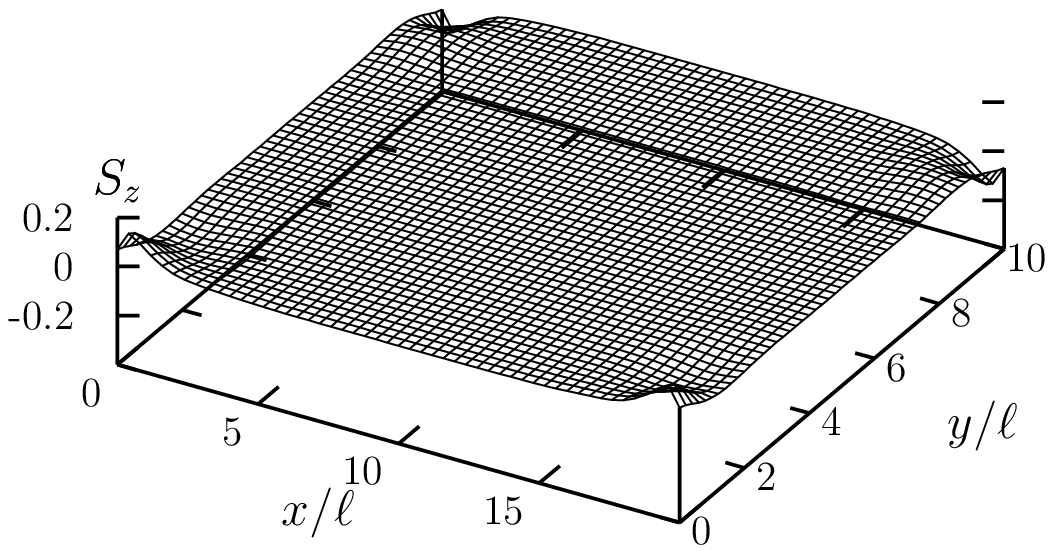} 
\caption{
\label{Fig1}
Spin polarization in the presence of an
electrical current flowing in $x$-direction for a strip of length
$L_x=20 l $ and $L_y=10 l$. The spin-orbit coupling strength is $\alpha
=10^{-3}v_F$ and the elastic scattering rate is $1/\tau = \alpha p_F/ 2  $.
The spin polarization is given in units of the bulk value, $S_0 = -{|e| E \alpha \tau
N_0}$.} 
\end{figure}
First we show numerical results for the spin polarization in the
stationary limit.
Fig.~\ref{Fig1} depicts the voltage induced spin polarization 
for 
$L_x = 20l$, $L_y = 10l$, $ \alpha p_F  \tau = 2 $ and $\alpha/v_F =
10^{-3}$; all our results are linear in the applied voltage, due to
the linearity of the underlying equations.
In the bulk, only the $S_y$ component is nonzero, and given by
$S_0 = - | e|  E \alpha \tau N_0  $
[\onlinecite{edelstein1990,mishchenko2004}].
A spin-Hall effect induced spin polarization is found in the corners,
as it is expected in [\onlinecite{mishchenko2004}]. The spin
polarization however is not purely in $z$-direction but has also
components in $x$-direction. 

\begin{figure}
\includegraphics[width=0.48\textwidth]{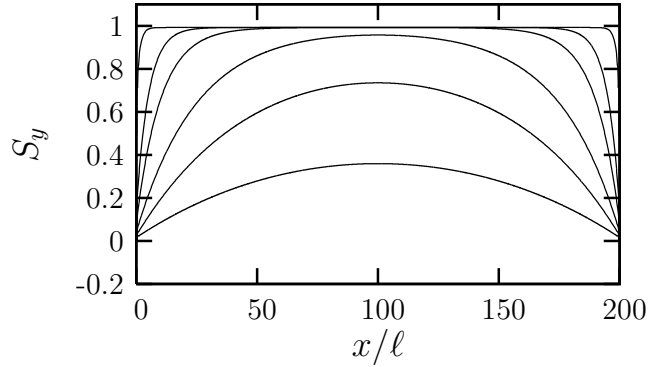} 
\caption{\label{Fig2}$S_y$ in units of $S_0$ as a function of 
$x$ for $L_x = 200l$, $L_y = 100l $, $\alpha/v_F =10^{-3}$ 
and $\alpha p_F \tau = 0.005, 0.01, 0.02, 0.05,0.1, 1$ (from bottom to
top). }
\end{figure}
Fig.~\ref{Fig2} shows $S_y(x, y\! = \! L_y/2 )$ by varying disorder.
In the diffusive limit and assuming that the spin polarization
vanishes at the interface to the leads, it has been predicted
that [\onlinecite{mishchenko2004}] 
\begin{equation}
S_y (x) = S_0      
\left( 1- \frac{\cosh[(x-L_x/2)/L_s]}{\cosh(L_x/2L_s)} \right)
, \end{equation}
where $L_s$ is the spin relaxation length.
Apparently with the boundary condition we choose a spin
polarization still exists near $x=0$, $L_x$, in particular in the clean
limit.
Some mean free paths away from the interface on the other
hand the data can be well fitted with an exponential increase or
decrease, both in the clean and dirty limit. 
\begin{figure}
\includegraphics[width=0.48\textwidth]{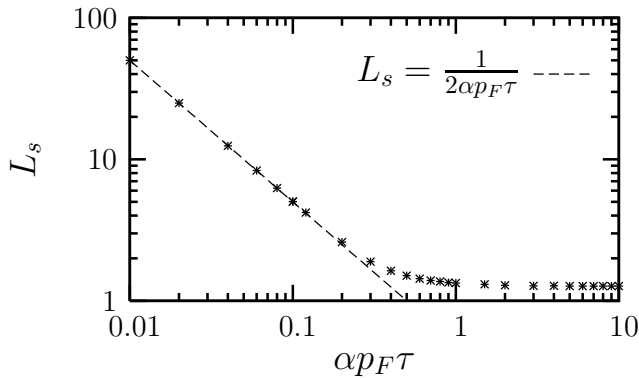} 
\caption{\label{Fig3}Spin relaxation length $L_s$ in units of $l$ as a function of disorder,
obtained by fitting the spatial dependence of the electric field induced spin
polarization (shown in Fig.~\ref{Fig2}) using 
$ S_y = a + b [ \exp(-x/L_s) + \exp(-(L_x - x)/L_s ) ] $.
The diffusive limit expression is shown as a dashed line.}
\end{figure}
As a result we obtain the spin relaxation length as a function of
disorder, shown in Fig.~\ref{Fig3}. In the dirty limit, 
$\alpha p_F \tau \ll 1 $, our numerical result agrees with what is expected from 
the diffusion equation, 
$L_s = \sqrt{D \tau_s} = l / 2 \alpha p_F \tau$. 
In the clean limit, for which we are not aware of any quantitative
predictions, the spin relaxation length is of the order of the
mean free path, $L_s \approx 1.27l$.

In the following we will consider the time evolution of the
spin polarization and the spin current.
We start with a system in thermal equilibrium, switch on
the voltage and observe the relaxation of the system into its stationary
non-equilibrium state.
It is a nontrivial problem to describe such a situation theoretically.
One might be tempted to
allow a time dependent voltage in the boundary
condition, Eq.~(\ref{eq50}),
and to follow then the time evolution.
In this case the charge density becomes time dependent and
inhomogeneous. This procedure makes sense for non-interacting
electrons, but not for interacting electrons where the long range
Coulomb interaction enforces charge neutrality.
In principle, the interaction can be included into the quasiclassical 
formalism explicitly, see e.g.~\cite{schwab2003}, however this is beyond the scope
of the present paper.

Instead, we assume in the following that a voltage difference across the
leads instantly results in a homogeneous electric field in the sample.
Thus one has to solve Eq.~(\ref{eqEilenberger}) 
with the initial condition
$g({\bf p}, {\epsilon}; {\bf x}, t) = \tanh(\epsilon/2T)( g^R - g^A )$ 
and taking into account the electric field
via the substitution $\partial_{\bf x} \to \partial_{\bf x }- |e| {\bf E} \partial_\epsilon$. 
In the numerics, however we find it
more convenient to work in a scalar gauge, since then the (static)
electric field disappears from the equation-of-motion and is present
only in the initial condition and in the boundary condition.
Generally the gauge transformation for the fields and the Green
function reads 
\begin{eqnarray}
{ \bf A }  &\to & {\bf A } + \partial_{\bf x} \chi \\
{ \phi  }  & \to & \phi - \partial_t  \chi \\
g_{t_1 t_2} & \to &\exp\{-{\rm i} |e| [\chi(t_1) - \chi(t_2)]  \} g_{t_1 t_2} 
.\end{eqnarray}
In the end we have to solve Eq.~(\ref{eqEilenberger}) with the
boundary condition (\ref{eq50})
and the initial condition 
\begin{equation}
g^K( \hat {\bf p}, {\epsilon}; {\bf x}, t=0)= \tanh\left(\frac{\epsilon +
|e| \phi(x)}{2 T }\right) (g^R-g^A),
\end{equation}
where $\phi(x)$ interpolates linearly between the two leads, 
$\phi(x) = V (L_x/2-x)/L_x$.

\begin{figure}
 \centerline{\includegraphics[width=0.48\textwidth]{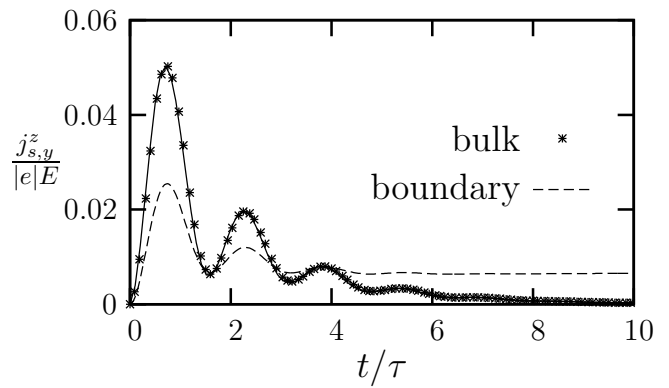}}
 \caption{
 \label{fig5}
 Time evolution of the spin-Hall current at the interface
 to the leads and in the bulk. In the bulk we compare our numerical
 result (data points) with the analytical result (full line) of Eq.\ (\ref{jz_analytical}).
 Near the leads, only numerical data are available (dashed curve). 
 $j_{s,y}^z$ is evaluated at $y =  L_y/2$, $x=0$ (boundary) and $x=
 L_x/2$ (bulk) for
 $L_x = 20 l$, $L_y = 10 l $, $\alpha/v_F =10^{-3}$ and $
 \alpha p_F \tau = 2$.
 }
\end{figure}
In Fig.~(\ref{fig5}) we show the spin current $j^z_{s,y}$ as a function of time in the bulk
and at the interface to the leads of a rather clean system ($ \alpha p_F  \tau = 2$,
$\alpha/ v_F = 10^{-3}$). On short time scales the bulk current agrees with what
we found ignoring disorder in Sect.~\ref{CleanLimit}: The spin current oscillates as
a function of time with frequency $2 \alpha p_F$, the time average is
given by the universal spin-Hall conductivity. 
In the bulk, for the weakly disordered system we are considering, the time-dependent 
spin-current is given by 
\begin{equation}
\label{jz_analytical}
j_{s,y}^z = \frac{ |e|  E}{8\pi} 
\left[ \exp(-t/2 \tau ) - \exp(-3t/4\tau) \cos(2\alpha p_F t ) \right]
,\end{equation}
which can be obtained from the frequency dependent spin-Hall
conductivity given in the appendix.
On the time scale of the spin relaxation time, here given by the
scattering time $\tau$, the bulk spin current becomes exponentially
suppressed and goes to zero in the stationary limit.
Near the leads, on the other hand, the situation is somewhat different,
since a finite spin current remains in the stationary limit.

\begin{figure}
 \centerline{\includegraphics[width=0.48\textwidth]{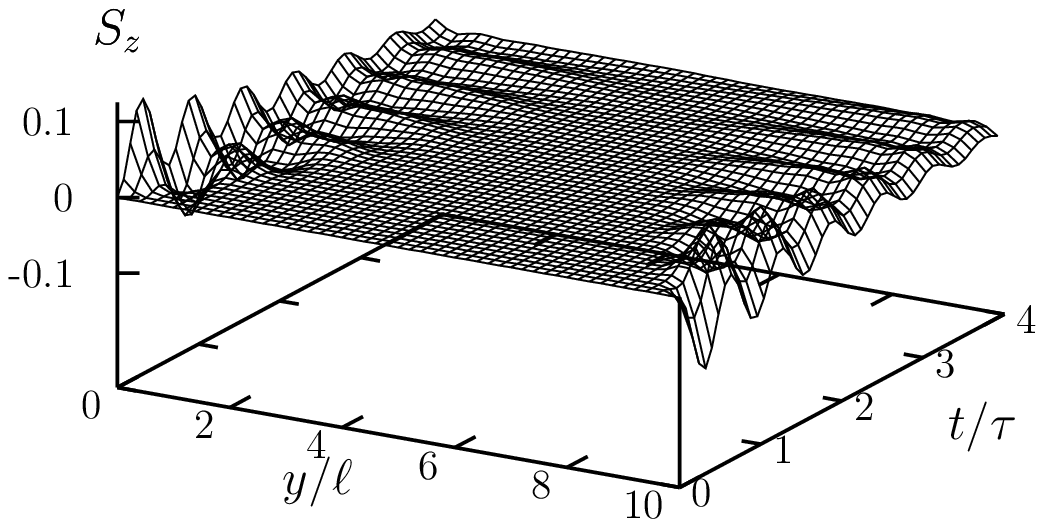}}
 \centerline{\includegraphics[width=0.48\textwidth]{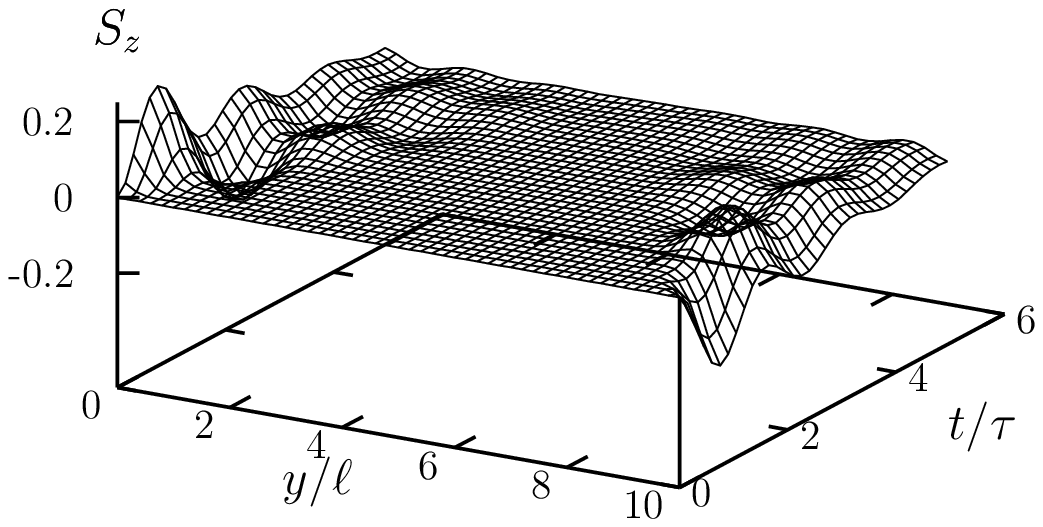}}
 \centerline{\includegraphics[width=0.48\textwidth]{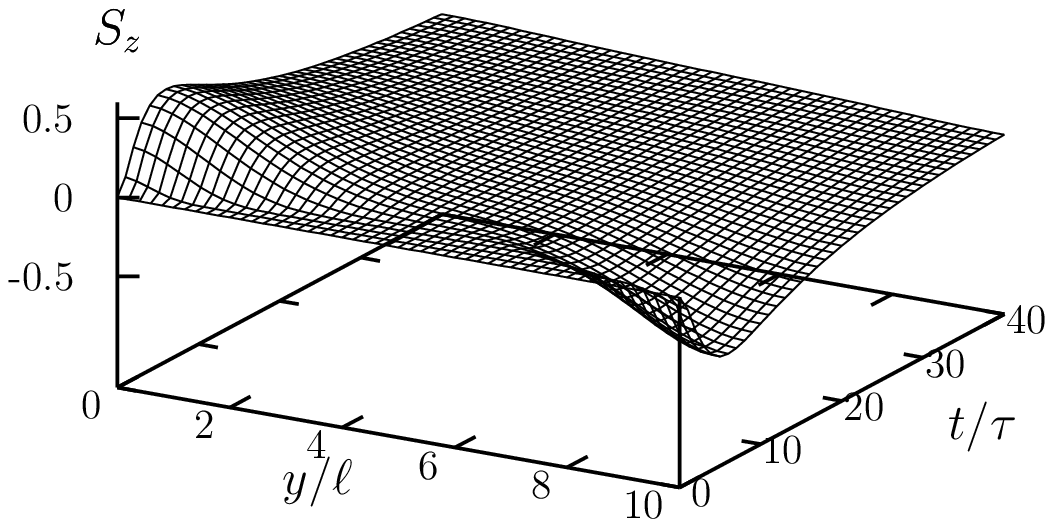}}
 \caption{Spin-Hall effect induced spin polarization $S_z$ in units 
 of $S_0$
 as a function of $y$ and $t$ at 
 $x= L_x/2$ for $L_x = 20 l$, $L_y =10 l$, $\alpha/v_F = 10^{-3}$, and
 $\alpha p_F \tau = 0.25, 2, 5$ (from bottom to top). 
}
\label{fig6}
\end{figure}

An important question is whether the spin current polarizes the
electron system  
at the edges. In Fig.~\ref{fig6} we show the spin polarization in 
$z$-direction across the system at $x=L_x/2$ as a function of time.
Since in the early time evolution spin current flows in the bulk, spin
density accumulates near the edges. When the spin current disappears
also the polarization vanishes. We see that the spin polarization at the edges oscillates
as expected with frequency $2\alpha p_F$. In the cleaner systems oscillations are of course faster.
Remarkably, the maximum amplitude of oscillation relative to the bulk
value
is larger in the dirty system ($\alpha p_F \tau =0.25$ ), 
where it is almost of  the order of one. 

This can be understood as follows: a rough estimate of the spin
polarization at the edge is $s^z \sim \tau_s j_{s,y}^z /L_s$. With
\begin{eqnarray}
\tau_s & \sim & \tau/(\alpha p_F \tau)^2 \\
 j_{s,y}^z  &\sim & e E (\alpha p_F \tau)^2 \\
 L_s & \sim & l/(\alpha p_F \tau)
 \end{eqnarray}
the result is indeed
$s^z \sim S_0  = e E \alpha \tau N_0$.
In the clean limit, on the other hand, the typical time and length
scales are $\tau_s \sim \tau$ and $L_s \sim l$, from which we estimate 
$s^z \sim S_0 /(\alpha p_F \tau)$, in agreement with our numerical
findings.

We expect the detailed structure of the spin polarization near
the edges to depend on the boundary conditions, 
i.e., on the type of the confinement at the edges. How boundary conditions affect the spin polarization at the edges
is certainly an experimentally relevant issue, which however is beyond the scope of this paper. 

\section{Summary}
We studied the spin-Hall effect in a two-dimensional electron system
by applying the method of quasiclassical Green's functions. The method
has its strength in the description of macroscopic systems, i.e. systems
with linear dimensions that are large compared to the Fermi
wavelength. In particular we derived an Eilenberger-like equation in the presence of
a generic spin-orbit coupling. 
We also showed that the method allows to derive in an elegant way
various results present in the literature.
For the special case of the Rashba model
we calculated numerically the spin-Hall current and the spin-Hall effect induced
spin polarization on a strip. 
From our data we were able to extract quantitatively the spin relaxation length 
in the entire regime from the clean to the dirty limit,
which is not covered by the diffusion equation approach.
Although in the Rashba model the zero-frequency spin-Hall
conductivity is zero we found a spin-Hall induced spin polarization
at the edges on a short time scale after switching on the voltage.
We believe that our approach will stimulate further work in the field.

We thank U.\ Eckern for valuable discussions. 
This work was supported by the Deutsche Forschungsgemeinschaft through
SFB 484.
\appendix
\section{Diffusive limit}
When spatial and temporal variations are slow
it is in many cases convenient to study the equations of motion in the diffusive limit.
Usually, in this limit the full angular dependent Green function 
${\check g}(\hat {\bf p})$ can be easily constructed from its angular
average $\langle {\check g} (\hat {\bf p}) \rangle$ so it is sufficient to
study $\langle {\check g} (\hat {\bf p}) \rangle$.
To determine the equation for $\langle {\check g} (\hat {\bf p}) \rangle$  we first write 
all the terms of Eq.\ (\ref{eqEilenberger}) with ${\check g}$ on the left hand
side and those depending on  $\langle {\check g} \rangle $ on the right
hand side.  For the Keldysh component we hence  get
\begin{equation} \label{eqA1}
( M_0 + M_1  ) g^K =  (N_0 + N_1 ) \langle g^K \rangle
\end{equation}
where 
\begin{eqnarray}
M_0 g^K  & = & g^K + \tau \partial_t g^K 
  + v_F \tau \hat{ \bf p} \cdot \partial_{\bf x } g^K \nonumber \\
  & +& {\rm i } \tau  \left[ {\bf b}_0 \cdot {\boldsymbol \sigma } ,  g^K \right] \\
M_1 g^K & = & -\frac{1}{2}  \tau
  \left\{ 
  \frac{ ({ \bf  b}_0 \cdot {\boldsymbol \sigma }) \hat{ \bf p } }{ p_F }  
  - \partial_{\bf p }({\bf b}_0 \cdot {\boldsymbol \sigma} ) ,
    \partial_{\bf x } g^K \right\} \cr
&& - \frac{1}{2} {\rm i } \tau 
     \left[ \partial_\xi ({\bf b}_0 \cdot {\boldsymbol \sigma }),
     \left\{{\bf b}_0 \cdot {\boldsymbol \sigma }  ,  g^K \right\}
     \right ]  \cr
&& - \frac{1}{2} \left\{ 
     \langle \partial_\xi  {\bf b}_0 \cdot {\boldsymbol  \sigma  } \rangle  , g^K \right\}  \\
N_0  \langle g^K \rangle  &=  & \langle g^K \rangle  \\
N_1  \langle g^K \rangle  & = &- \frac{1}{2} 
          \left\{ \partial_\xi {\bf b}_0 \cdot {\boldsymbol \sigma  }, \langle  g^K  \rangle \right\}
.\end{eqnarray}
Here $M_1$ and $N_1$ are small in the expansion parameter
$ |{\bf b}|/\epsilon_F$.
The Eilenberger equation is then rewritten as
\begin{equation}
{ g^K} =(M_0 + M_1)^{-1}(N_0 + N_1) \langle { g^K} \rangle 
,\end{equation}
i.e.\ to first order in $|{\bf b}|/\epsilon_F $
\begin{equation}
{ g^K} = \left( M_0^{-1} + M_0^{-1}N_1 - M_0^{-1} M_1 M_0^{-1} \right) 
 \langle g^K \rangle 
,\end{equation}
from which the equation for the $s$-wave component of the Green
function becomes
\begin{equation}
\left( 1-\langle M_0^{-1} \rangle  - \langle M_0^{-1}N_1 \rangle  +
\langle M_0^{-1} M_1 M_0^{-1} \rangle \right)  \langle { g^K} \rangle =0
.\end{equation} 
In the low frequency, long wavelength limit
this is the generalized diffusion equation obtained in the
literature in various limits
\cite{mishchenko2004,burkov2004,malshukov2005}.
The explicit form is obtained by evaluating the angular
average of the operator product ${ M}^{-1}N$.

In the Rashba model for instance where 
${\bf b} = \alpha   {\bf p } \times \hat{\bf e}_z  $ 
one finds
\begin{eqnarray}
M_0 & =  & \left(
\begin{array}{cccc}
     L       &   0   &   0   & 0 \\
     0       &   L   &   0   & a \hat p_x \\
     0       &   0   &   L   & a\hat p_y \\
     0       & -a  \hat p_x &
               -a  \hat p_y & L
\end{array}
\right) \\[1mm]
M_1 & = &\left(
\begin{array}{cccc}
     0       &    Q_y & - Q_x & 0  \\
 Q_y  &   0            &   0          &  0  \\
-Q_x  &   0            &   0          &   0  \\
     0       &   0            &   0          &  0 
\end{array}
\right)
\end{eqnarray}
with
\begin{eqnarray}
L & = & 1 + \tau \partial_t + v_F \tau \hat {\bf p } \cdot \partial_{\bf x } \\
a &= & 2 \alpha p_F \tau  \\
Q_{x,y} &= & \alpha \tau (\partial_{x,y}-(\pup \cdot
\partial_{\bf x} ) {\hat p}_{x,y}) 
\end{eqnarray}
and
\begin{equation}
N_0+ N_1 = \left(
\begin{array}{cccc}
   1    & -{\alpha }\hat{ p }_y /v_F & {\alpha}\hat{p}_x /v_F & 0 \\
  -{\alpha} \hat{p}_y/v_F &   1    &   0 &  0 \\
   {\alpha} \hat{p}_x/v_F &   0    &   1 &  0 \\
   0   &   0    &  0  & 1
\end{array}
\right)
.\end{equation}
In a dirty system, where $a \ll 1 $,
the result reads
\begin{widetext}
\begin{equation}
\label{eqA15}
\left(
\begin{array}{cccc}
\partial_t -  D \partial_{\bf x}^2  & -2  B \partial_y &2 B \partial_x &     0 \\
-2 B \partial_y        &  \partial_t -D \partial_{\bf x}^2+\tau_s^{-1}& 0      &-2 C \partial_x \\
2B \partial_x & 0 & \partial_t - D \partial_{\bf x}^2+\tau_s^{-1}&-2 C \partial_y \\
0&2C\partial_x &2 C \partial_y & \partial_t  - D \partial_{\bf x}^2+2\tau_s^{-1}\\
\end{array}
\right)
 \left(
  \begin{array}{c}
    \langle g_0^K \rangle \\ 
    \langle g_x^K \rangle \\ 
    \langle g_y^K \rangle \\ 
    \langle g_z^K \rangle \\ 
  \end{array}
  \right) =0
\end{equation}
\end{widetext}
where $D= \frac{1}{2} v_F^2 \tau$ is the diffusion constant and
\begin{equation}
B=\frac{\alpha a^2}{1+a^2}, \   \   C=\frac{v_F a}{2(1+a^2)^2}, \   \
\frac{1}{\tau_s}=\frac{1}{2\tau}\frac{a^2}{1+a^2}
.\end{equation}
In the clean limit ($ a \stackrel{>}{\sim}1  $) there is no spin diffusion,
and Eq.~(\ref{eqA15}) is not justified. However the
equation is constructed in such a way  that it applies in the clean
limit for a space and time independent spin polarization.

In the general case (arbitrary $a$) we
investigate spin relaxation
for a spatially homogeneous system, where the equation of motion reads
\begin{eqnarray}
\left[ L(L^2 + a^2) -L^2 -\frac{1}{2} a^2 \right] \langle g_{x,y}^K \rangle  &= & 0 \\
\left[ L(L^2 + a^2) -L^2                  \right] \langle g_{z  }^K \rangle  &= & 0 
\end{eqnarray}
with $L = 1 + \tau \partial_t$.
Clearly the spin-dynamics for the 
each component is determined from three relaxation times. For the 
$z$-component, e.g., these are
\begin{eqnarray}
\frac{1}{\tau_1}     &= &  \frac{1}{\tau } \\
\frac{1}{\tau_{2,3}} &= &  \frac{1}{2 \tau } \pm \frac{1}{2 \tau}
\sqrt{1-4a^2}.
\end{eqnarray}
For dirty systems the longest of these
times is $\tau_3 \approx \tau /a^2$, in agreement with what we
find from the diffusion equation (\ref{eqA15}). In the clean limit the
``relaxation'' time  becomes complex corresponding to an
oscillating spin polarization. 

\section{Spin-Hall conductivity}

We go back to (\ref{eqA1}) and solve it for an infinite
system under the influence of a uniform but time-dependent electric field
in  $x$-direction.
We choose the vector gauge as in section \ref{CleanLimit}, i.e.
$\partial_{\bf x } \to \partial_{\bf x} -|e| E \partial_\epsilon$.
To linear order in the external field and transforming  to Fourier space,
(\ref{eqA1}) becomes:
\begin{equation}
{ M}_{0 }g^K  = (1 + N_1) \langle g^K \rangle + S_E 
,\end{equation}
where $M_0$ and $N_1$ are the same as before, $S_E$ is a source
term due to the electric field:
\begin{equation}
S_E = \tilde{E}
\left[
\left(
\begin{array}{c}
\hat{ p}_x \\
0 \\
0 \\
0 \\
\end{array}
\right)
+  \frac{\alpha}{v_F}
\left(
\begin{array}{c}
0 \\
-2 \hat{ p}_x \hat{ p}_y \\
\hat{ p}_x^2 - \hat{ p}_y^2 \\
0 \\
\end{array}
\right)
\right]
\end{equation}
with  $\tilde{E}=|e|lE(\omega)\partial_{\epsilon}g_0^{K,\rm eq}$.
Inverting ${ M}_{0}$ and performing the angular average one obtains
\begin{equation}
\langle \hat{ p}_y g_z^K \rangle  = -\frac{a}{2(L_{}^2 + a^2)}
\left[ \frac{\alpha}{v_F} \tilde{E}
 -\langle g_y^K \rangle
\right]\label{zspincurrent}
\end{equation}
where the two terms in square brackets correspond to {\sl bubble} and {\sl vertex} corrections, respectively,
in the diagrammatic language. Furthermore we see
that 
the spin polarization along  ${ {\bf e}_y}$
contributes to the spin-Hall current. The spin polarization along ${ {\bf e}_y}$
is obtained by inverting ${ M}_{0}$ and performing the angular average
\begin{equation}
\langle g_y^K \rangle = \frac{\alpha }{v_F} \frac{a^2\tilde{E}}{2\left[L_{}(L_{}^2 + a^2)-
\left(L_{}^2 + \frac{a^2}{2}\right)\right]}.\label{yspinpolarization}
\end{equation}
In the static limit ($L=1$) this reads $ \langle g_y^K \rangle =
\frac{\alpha }{v_F} \tilde E $, corresponding to the bulk value
$S_0$, quoted in the text. This value for $ \langle g_y^K \rangle $ leads to the cancellation discovered 
previously \cite{schwab2002} and to the vanishing of the static spin Hall conductivity.
Finally, by combining Eqs.~(\ref{zspincurrent}-\ref{yspinpolarization}),
 one  obtains the frequency-dependent spin-Hall current  with polarization along ${ {\bf e}_z}$ as
\begin{eqnarray}
j^z_{s,y}(\omega)& =&\sigma_{sH}(\omega )E(\omega )\nonumber\\
&=& \frac{|e|}{8\pi}\frac{-i\omega\tau a^2}{L_{}(L_{}^2 + a^2)-
\left(L_{}^2 + \frac{a^2}{2}\right)}E(\omega)
\end{eqnarray}
which agrees with the result given by both Mishchenko et al.
\cite{mishchenko2004} and Chalaev and Loss \cite{chalaev2005}. 
To examine the transient response in time, we use
 $E(t)=E\theta (t)$ so that
 \begin{equation}
j^z_{s,y}(t) =E \int_{-\infty}^t d t' \sigma_{sH}(t')
\label{b7}
\end{equation}
where 
\begin{equation}
\sigma_{sH}(t)=\int_{-\infty}^{\infty}\frac{d \omega}{2\pi}\sigma_{sH}(\omega )
e^{-{\rm i}\omega t}.
\end{equation}
The poles of the integrand may be found by expressing $\omega\tau =x+ {\rm i}y$ and setting to zero the real and imaginary part of the cubic polynomial in the denominator of $\sigma_{sH}(\omega )$
\begin{eqnarray}
x[x^2-3y^2-4y-(1+a^2)]&=&0 \\
x^2(3y+2)-y^3-2y^2-y(1+a^2)-\frac{a^2}{2}&=&0.
\end{eqnarray}
In the large $a$ limit, one obtains the three solutions 
\begin{equation}
x=0, \  y=-1/2; \   \    \  x=\pm a, \ y=-3/4.
\end{equation}
Computing the residues with the same accuracy yields
\begin{eqnarray}
\sigma_{sH}(t)&= &-\frac{|e|}{8\pi}\frac{\theta(t)}{2\tau} \big[ 
{\rm e}^{-t/2\tau} 
 -{\rm e }^{-3t/4\tau} \cos(a t/\tau) 
\nonumber \\[1mm] 
&&  
-2 a    {\rm e }^{-3t/4\tau} \sin(a t/\tau)     \big].
\end{eqnarray}
Finally, inserting the above result into Eq.~(\ref{b7}), one obtains
-- to leading order in $1/a$ -- the time-dependent current quoted in the text.


\begin{thebibliography}{999}
\bibitem{dyakonov1971}M.\ I.\ Dyakonov and V.\ I.\ Perel, 
                Phys.\ Lett.\ {\bf 35A}, 459 (1971).
\bibitem{edelstein1990}V.\ M.\ Edelstein, 
                Sol.\ Stat.\ Commun.\ {\bf 73}, 233 (1990);
                J.\ Phys.: Condens.\ Matter {\bf 5}, 2603 (1993). 
\bibitem{hirsch1999}J.\ E.\ Hirsch, Phys.\ Rev.\ Lett.\ {\bf 83}, 1834 (1999).
\bibitem{murakami2003} S.\ Murakami, N.\ Nagaosa, and S.-C.\ Zhang, Science {\bf  301}, 1348 (2003).
\bibitem{sinova2004} J.\ Sinova, D.\ Culcer, Q.\ Niu, N.\ A.\ Sinitsyn, T.\ Jungwirth, and A.\ H.\ MacDonald,
                Phys.\ Rev.\ Lett.\ {\bf 92}, 126603 (2004).
\bibitem{kato2004}Y.\ K.\ Kato {\em et al.}, Science {\bf 306}, 1910 (2004).
\bibitem{wunderlich2005}J.\ Wunderlich,  B.\  Kaestner,  J. \ Sinova,  and T. \ Jungwirth, 
Phys.\ Rev.\ Lett.\ {\bf 94}, 047204 (2005).
\bibitem{sinitsyn2004}N.\ A.\ Sinitsyn, E.\ M.\ Hankiewicz, W.\ Teizer, and J.\ Sinova,
                Phys.\ Rev.\ B {\bf 70}, 081312(R)  (2004).
\bibitem{shen2004}S.-Q.\ Shen, Phys.\ Rev.\ B {\bf 70}, 081311(R)  (2004). 
\bibitem{chang2005}M.-C.\ Chang, Phys.\ Rev.\ B {\bf 71}, 085315 (2005).
\bibitem{shytov2005} A.\ V.\ Shytov, E.\ G.\ Mishchenko, H.-A.\ Engel, and B.\ I.\ Halperin, cond-mat/0509702.
\bibitem{inoue2004}J.\ I. \ Inoue, G.\ E.\ W.\ Bauer, and L.\ W.\ Molenkamp, 
                Phys.\ Rev.\ B {\bf 70}, 041303 (2004).
\bibitem{mishchenko2004} E.\ G.\ Mishchenko, A.\ V.\ Shytov, and B.\ I.\ Halperin, 
                Phys.\ Rev.\ Lett.\  {\bf 93}, 226602 (2004).
\bibitem{khaetskii2004}A.\ Khaetskii,
                cond-mat/0408136.
\bibitem{raimondi2005} R.\ Raimondi and P.\ Schwab, 
                Phys.\ Rev.\ B {\bf 71}, 033311 (2005).
\bibitem{dimitrova2005}O.\ V.\ Dimitrova, 
                Phys. Rev. B {\bf 71}, 245327 (2005).
\bibitem{chalaev2005}O.\ Chalaev and D.\ Loss, Phys.\ Rev.\ B {\bf 71}, 245318 (2005). 
\bibitem{nomura2005} K.\ Nomura, J.\ Sinova, N.\ A.\ Sinitsyn, and A.\ H.\ MacDonald,
                Phys.\ Rev.\ B {\bf 72}, 165316 (2005).
\bibitem{rashba2004} E.\ I.\ Rashba,
                Phys.\ Rev.\ B {\bf 70}, 201309(R)  (2004).
\bibitem{murakami2004}S.\ Murakami, Phys.\ Rev.\ B {\bf 69}, 241202(R) (2004).
\bibitem{malshukov2005} A.\ G.\ Mal'shukov, L.\ Y.\ Wang, C.\ S.\ Chu, and K.\ A.\ Chao, 
                Phys.\ Rev.\ Lett.\ {\bf 95}, 146601 (2005).
\bibitem{khaetskii2005}A.\ Khaetskii,
                cond-mat/0510815.
\bibitem{nikolic2005}B.\ K.\ Nikolic, S.\ Souma, L.\ P.\ Zarbo, and J.\ Sinova,
                Phys.\ Rev.\ Lett.\ {\bf 95}, 046601 (2005).
\bibitem{onada2005}M.\ Onoda and N.\ Nagaosa,
                Phys.\ Rev.\ B {\bf 72}, 081301(R) (2005).
\bibitem{burkov2004}A.\ A.\ Burkov, A.\ S.\ Nunez, and A.\ H.\ MacDonald,
                Phys.\ Rev.\ B {\bf 70}, 155308 (2004).
\bibitem{rammer1986}J.\ Rammer and H.\ Smith, Rev.\ Mod.\ Phys.\ {\bf
58}, 323 (1986).
\bibitem{zaitsev1984}A.\ V.\ Zaitsev,
         Zh.\ Eksp.\ Teor.\ Fiz.\ {\bf 86}, 1742 (1984) [Sov.\ Phys.-JETP {\bf 59}, 1015 (1984)].
\bibitem{shelankov1985} A.\ L.\ Shelankov, J.\ Low Temp.\ Phys.\ {\bf 60}, 29 (1985).
\bibitem{millis1988} A.\ Millis, D.\ Rainer, and J.\ A.\ Sauls, Phys.\ Rev.\ B
         {\bf 38}, 4504 (1988).

\bibitem{silvestrov2005}P.\ G.\ Silvestrov and E.\ G.\ Mishchenko, cond-mat/0506516

\bibitem{schwab2003}P.\ Schwab and R.\ Raimondi, 
                Ann.\ Phys.\ (Leipzig) {\bf 12}, 471 (2003).
\bibitem{schwab2002} P.\ Schwab and R.\ Raimondi, 
                Eur.\ Phys.\ J. B {\bf 25}, 483 (2002).


\end{thebibliography}
\end{document}